\documentclass[sigconf]{acmart}
\usepackage{booktabs} 

\setcopyright{rightsretained}

\begin{document}

\copyrightyear{2019}
\acmYear{2019}
\acmConference[CECC 2019]{Central European Cybersecurity Conference}{November 14--15, 2019}{Munich, Germany}
\acmBooktitle{Central European Cybersecurity Conference (CECC 2019), November 14--15, 2019, Munich, Germany}
\acmPrice{15.00}
\acmDOI{10.1145/3360664.3362698}
\acmISBN{978-1-4503-7296-1/19/11}

\title{Integrating Threat Modeling and Automated Test Case Generation into Industrialized Software Security Testing}

\author{Stefan Marksteiner}
\email{stefan.marksteiner@avl.com}
\orcid{0000-0001-8556-1541}
\affiliation{%
  \institution{AVL List GmbH}
  \city{Graz}
  \country{Austria}
}
\author{Rudolf Ramler}
\email{rudolf.ramler@scch.at}
\orcid{0000-0001-9903-6107}
\affiliation{%
  \institution{Software Competence Center Hagenberg GmbH}
  \city{Hagenberg}
  \country{Austria}
}
\author{Hannes Sochor}
\email{hannes.sochor@scch.at}
\orcid{0000-0001-9903-6107}
\affiliation{%
  \institution{Software Competence Center Hagenberg GmbH}
  \city{Hagenberg}
  \country{Austria}
}

\renewcommand{\shortauthors}{Marksteiner, Ramler, Sochor}

\renewcommand{\shorttitle}{Integrating Threat Modeling and Automated Test Case Generation}

\begin{abstract}
Industrial Internet of Things (IIoT) application provide a whole new set of possibilities to drive efficiency of industrial production forward. However, with the higher degree of integration among systems, 
comes a plethora of new threats to the latter, as they are not yet designed to be broadly reachable and interoperable. To mitigate these vast amount of new threats, systematic and automated test methods
are necessary. This comprehensiveness can be achieved by thorough threat modeling. In order to automate security test, we present an approach to automate the testing process from threat modeling onward, closing
the gap between threat modeling and automated test case generation.
\end{abstract}

%
%



\maketitle

\section{Introduction and Motivation}
\label{sec:intro}
The increasing connectivity of cyber-physical objects in critical environments like industrial production or autonomous driving leads to an increasing vulnerability of the overall IoT-based, cyber-physical systems (CPS). 
To ensure security in large, heterogeneous, distributed and dynamic environments, a clear understanding of the related properties and threats and how they are addressed during the entire system life cycle is necessary. 
Hence, such environments result in major challenges and a high effort for conventional approaches to quality assurance (QA) such as testing - still one of the most widely applied approaches in industry.
Conventional test and QA are characterized by manual tasks and script-based test automation, which limits the flexibility required for adapting to heterogeneous and dynamic environments~\cite{eberhardinger2014towards}.
Furthermore, adequately addressing security issues requires a deep integration with related activities and tools for model-driven security engineering.
This work presents following contributions in its next sections:
\begin{itemize}
\item Abstraction through system and threat modeling;
\item Feedback-directed test case generation (TCG) using attack patterns based on the threats derived from the model;
\item Automated execution of the generated security tests.
\end{itemize}

So far, only few practical efforts are known to automatically derive real-system executable test cases out of threat modeling (e.g.\cite{7102630}) where none of these uses a widely-known 
threat modeling tool.

\section{Threat Modeling and Test Case Generation}
\label{sec:TMTG}
\begin{figure}[h]
	\centerline{
		\scalebox{0.35}{
			\includegraphics{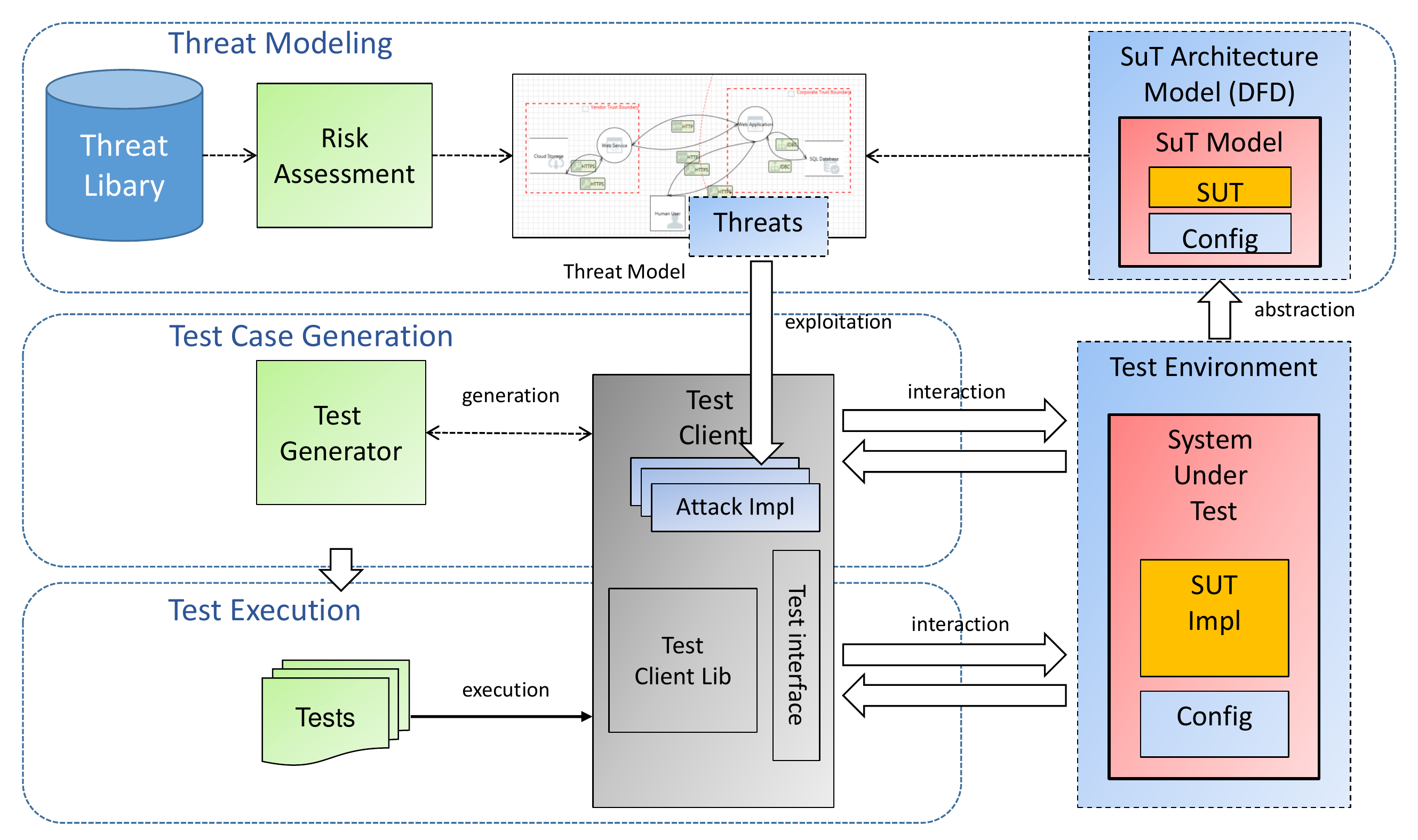}
		}
	}
\caption{Automated Testing Process}
\label{fig:arc}
\end{figure}

This section describes the components of threat-model driven TCG. Figure \ref{fig:arc} provides an overview.
\subsection{Threat Modeling}
In order to industrialize security testing, a model-based approach, abstracting the tests, is necessary for ensuring reusable test building blocks through the transaction from one test case to another.
Originally, Microsoft's Security Development Lifecycle contained threat modeling as a tool to identify security threats with the goal of creating a secure system design~\cite{shostack2014threat}. 
The concept has been expanded and transferred to various application domains and settings since~\cite{XIONG201953}. The approach uses data flow diagrams (DFDs), that contain all relevant system components of an SUT and to model how they interact with each other in terms of data exchange.
In order to derive relevant threats to the SUT, it uses threat templates that are usually derived by a commonly known methodology (e.g. STRIDE, VAST) and maps them to the DFD. This yields a list of threats 
that is both holistic and specific to the SUT. This can also be extended with security building blocks that refine the threat derivation~\cite{Sion:2018:SDF:3167132.3167285}.
The two basic components of a threat model therefore pose an ideal tool to achieve the system modeling for security TCG: the DFD represent an abstraction of the SUT, while the according 
threat template represent an abstraction of the  threats to the system. As the modeling combines these two into a single model, threats relevant to the SUT can be derived (in practice as XML output) and used to automatically generating attack scenarios as test cases.
In order to provide meaningful threats, the latter are preselected from a threat library and (as yet manually) run through a risk assessment matrix before being applied to the model. 

\subsection{Test Case Generation}

Automated security testing ranges from generating random input (fuzzing)~\cite{Takanen:2008:FSS:1404500} to security testing based on models~\cite{10.1002:stvr.1580}. 
The presented approach targets the generation of interaction sequences according to protocols used in IoT systems, the typical domain of MBT. However, instead of developing a 
full model, a list of commands (e.g., protocol messages) is specified and a test generator from functional testing is applied to create interaction sequences. Methods for generating sequences are 
based on random approaches, search-based strategies, genetic algorithms, reinforcement learning~\cite{ANAND20131978}.
Online feedback-directed testing is used to determine viable sequences and to avoid generating illegal sequences violating protocol constraints. This approach creates sequences by executing each 
selected command during test generation and evaluating the feedback (response of the SUT) before adding them to the sequence. Thus, the approach is able to generate long valid sequences (positive tests) 
that achieve a deep coverage of internal states as well as sequences that lead to an invalid command call after a number of valid interactions (negative tests). 
Security is addressed by augmenting the list of regular commands with the implementation of attack patterns obtained from publicly available catalogs like the 
Common Vulnerabilities and Exposures (CVE)~\cite{NIST:2011}, e.g. message flooding. 
Each attack pattern is implemented in the same way as a regular command call, making them accessible to the test generator. Hence, the test generator will typically create interaction sequences that 
consist of valid chunks of sequences interleaved by malicious interactions from attacks. 
The selection and prioritization of the attacks applied in test generation is controlled by the input provided by the threat model. An attack pattern is included if a corresponding vulnerability has 
been specified in the model.

\subsection{Automated Test Execution}
Test generation produces a set of executable test cases. Due to the online test generation approach, these test cases have already been executed in the generation step. Nevertheless, test execution is 
organized as a separate step to investigate any failing tests that have been generated but also to analyze the results produced by all other test (e.g. to evaluate specific attacks) while closely monitoring 
the SUT. 
Furthermore, by separating test generation and test execution, the generated tests can be executed multiple times as regression tests, in different test environments, and for varying configurations of the 
SUT~\cite{DANGLOT2019110398}.

\section{Results and Discussion}
\label{sec:Res}

A demonstrator for automated security testing has been built for the use case of secure data exchange for distributed connected devices in an untrusted environment. 
The approach is shown for data hubs using the MQTT messaging protocol as system under test (SUT). 
Security testing has been aligned with potential threats identified in a security risk assessment-based threat model, which, derived as described above, also poses one of this work's results.
The demonstrator has been developed to confirm the technical feasibility of the approach and to explore new opportunities for automation in the integration of threat modeling and security testing.

\section{Outlook}
\label{sec:Dis}
In this paper, we described the seamless integration of automated security testing for IoT systems with risk analysis-enhanced threat modeling.
The demonstrator, so far, has been used to confirm the principal technical feasibility of the approach. 
However, some practical problems need yet to be solved to provide a fully automated cybersecurity testing solution: on the one hand the 
template for the threat model needs to be extended with properties of modelled nodes (e.g. OS or network parameters) that would be needed by
the TCG execution to instantiate a proper SUT test environment as well as threat parameters. Secondly, there is significant 
engineering work to perform in order to actually implement attacks for the derived threats inside the TCG system. Also, the system so far lacks a driver for 
valid input. 
This is part of ongoing work, as well as 
 exploring further automation opportunities, in particular
a threat coverage analysis, differential and compatibility testing, as well as system and environment modeling.

\begin{acks}
This work was partly supported by the Austrian Research Promotion Agency (FFG) within the \textit{ICT of the future}
 grants program, grant nb. 863129 (project \textit{IoT4CPS}), of the Federal Ministry for Transport, Innovation and Technology
 (BMVIT).
\end{acks}

\bibliographystyle{ACM-Reference-Format}
\bibliography{literature} 

\end{document}